\documentclass{ws-procs9x6}
\usepackage{graphicx,picinpar,epsfig}

\begin{document}

\title{Constraints on the Decomposition of the Rotation Curves of Spiral
Galaxies}

\author{B. FUCHS}

\address{Astronomisches Rechen--Institut \\
M\"onchhofstr. 12--14 \\ 
69120 Heidelberg, Germany\\ 
E-mail: fuchs@ari.uni-heidelberg.de}


\maketitle

\abstracts{
I discuss anew how arguments about the internal dynamics of galactic disks set
constraints on the otherwise ambiguous decomposition of the rotation curves of 
spiral galaxies into the contributions by the various constituents of the 
galaxies. Analyzing the two sample galaxies NGC\,3198 and NGC\,2985 I conclude 
from the multiplicities of the spiral arms and the values of the $Q$ disk 
stability parameters that the disks of both galaxies are `maximum disks'.}

\section{Introduction}
The rotation curves of spiral galaxies provide the most direct evidence for the
presence of dark matter in galaxies. However, taken alone they do not
discriminate luminous from dark matter because their decomposition into the
contributions from the various constituents of the galaxies is highly ambiguous.
Thus further constraints are needed. Considerations of the dynamical state of
the resulting disk models can provide such constraints (Bosma 1999, Fuchs 1999).
Of particular interest is the question if the much discussed `maximum--disk'
models, i.e.~disks with their masses chosen at the maximum allowed by the data,
are dynamical viable disk models. The degeneracy of the decomposition problem 
is illustrated in Fig.~1 for the example of NGC\,3198. The rotation curve is
modelled as
\begin{equation}
v_{\rm c}^2(R) = v_{\rm c, disk}^2(R) + v_{\rm c, halo}^2(R) +
v_{\rm c, is\,gas}^2(R)\,,
\end{equation}
where $v_{\rm c, disk}$, $v_{\rm c, halo}$, and $v_{\rm c, is\,gas}$ denote
the contributions due to the stellar disk, the dark halo, and the interstellar
gas, respectively. The disk is modelled as an exponential disk and the dark halo
is described by a quasi--isothermal sphere. In the left panel of Fig.~1 the 
maximum--disk model of Broeils (1992) is reproduced, while the right panel
shows a submaximal disk model. The halo model parameters have been changed so
that both fits to the observed rotation curve are of the same quality.

\begin{center}
\hbox{
\epsfclipon
\epsfxsize=5.4cm
\epsffile{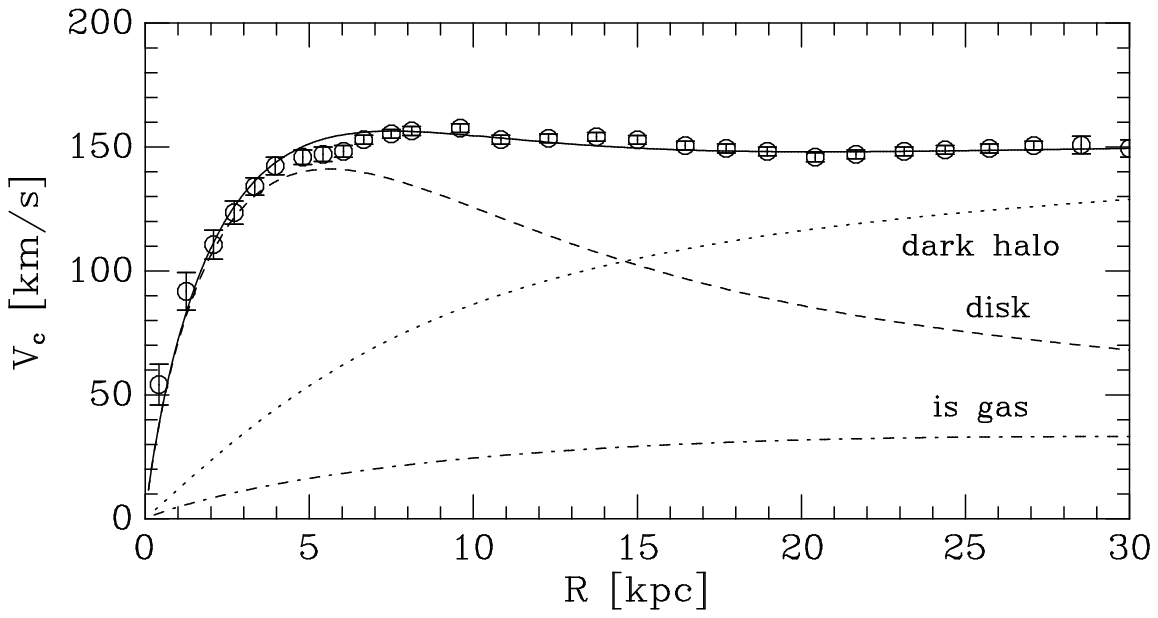}
\hspace{0.2cm}
\epsfclipon
\epsfxsize=5.4cm
\epsffile{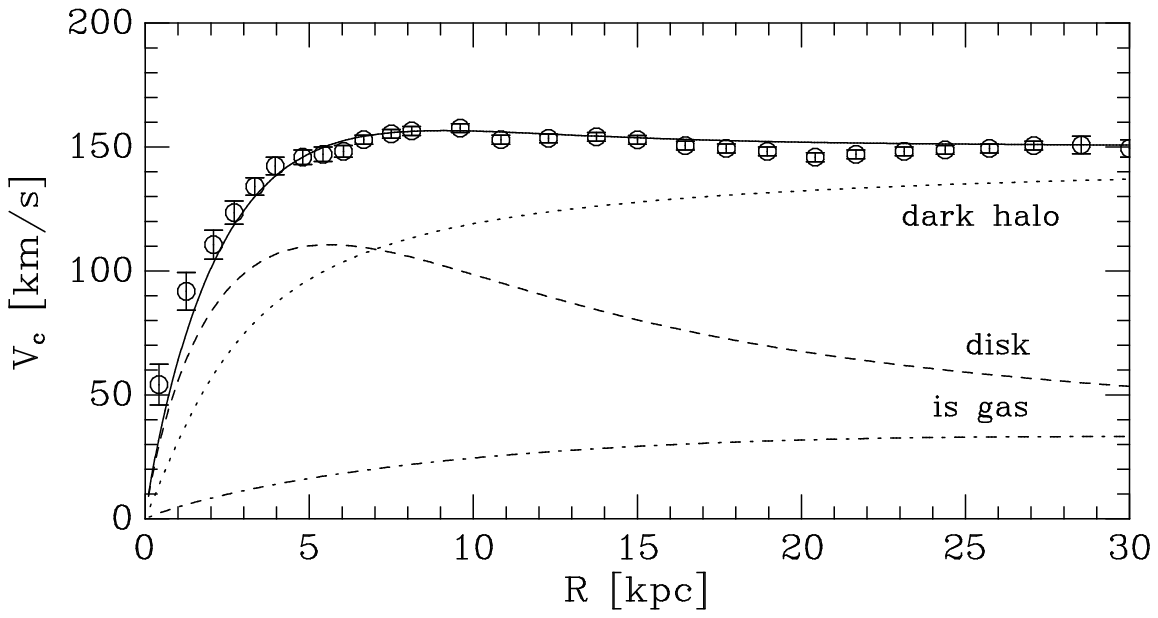}
}
\end{center}
\vspace{- 0.5cm} 
{\footnotesize Figure 1. Maximum and submaximal disk models of the rotation
curve of NGC\,3198.}

\section{Dynamical Constraints on the Decomposition of Rotation Curves}
The diagnostic tools I use to analyze the dynamical state of the disk models
are the Toomre stability parameter of the disks and, following Athanassoula et
al.~(1987), the predicted multiplicity of the spiral structures. The Toomre
stability parameter is given by
\begin{equation}
Q = \frac{\kappa \sigma_{\rm U}}{3.36 G \Sigma_{\rm d}}\,,
\end{equation}
where $\kappa$ denotes the epicyclic frequency, 
$\kappa = \sqrt{2}\,\frac{v_{\rm c}}{R}\,\sqrt{1+\frac{R}{v_{\rm c}}
\frac{d v_{\rm c}}{d R}}$.
$\sigma_{\rm U}$ is the radial velocity dispersion of the stars, $G$ the
constant of gravitation, and $\Sigma_{\rm d}$ the surface density of the disk.
The stability parameter must lie in the range 1 $< Q <$ 2, in order to prevent 
Jeans instability of the disk, on one hand, and
to allow the disks to develop spiral structures, on the other hand. If $Q$ were
less than 1, the disks would develop fierce dynamical instabilities and heat up
dynamically on very short time scales as demonstrated, for example, by Fuchs \&
von Linden (1998). Thus such disk models would be not equilibrium models.

The dynamics of galactic spiral structure is theoretically well understood in
the framework of the density wave theory of spiral arms. Density wave theory 
makes, in particular, a specific prediction for the number of spiral arms. 
Spiral density waves develop in galactic disks preferentially with a 
circumferential wavelength of about (Toomre 1981, Fuchs 2001, 2003c)
\begin{equation}
\lambda \approx X(\frac{A}{\Omega_0}) \lambda_{\rm crit} \,,
\end{equation}
where $\lambda_{\rm crit}$ denotes the critical wavelength
\begin{equation}
\lambda_{\rm crit} = \frac{4 \pi^2 G \Sigma_{\rm d}}{\kappa^2}\,.
\end{equation}
The coefficient $X(\frac{A}{\Omega_0})$
depends on the slope of the rotation curve measured by Oort's constant $A$,
$\frac{A}{\Omega_0} = \frac{1}{2} \left( 1 - \frac{R}{v_{\rm c}}
\frac{d v_{\rm c}}{d R} \right)$,
and has been determined explicitely for various cases by Toomre (1981),
Athanassoula (1984), or Fuchs (2001, 2003c). For a flat rotation curve the 
value is $X(0.5) = 2$. The number of spiral arms is obviously determined by how
often the wavelength $\lambda$ fits onto the annulus,
\begin{equation}
m \approx \frac{2 \pi R}{X \lambda_{\rm crit}}\,.
\end{equation}
The predicted number of spiral arms (5) is based on the local model of a
shearing sheet which describes the dynamics of a patch of a galactic disk
(Goldreich \& Lynden--Bell 1965, Julian \&Toomre 1966, Fuchs 2001). Equation (5)
can be applied globally to an entire disk strictly only in the case of a
Mestel disk, which has an exactly flat rotation curve and a surface density 
distribution falling radially off as $1/R$. In the exponential disk models 
used here $m$ varies formally with galactocentric distance.
However, the maximum growth factor of the amplitudes of the density
waves is not sharply peaked at the circumferential wavelength (3) (Toomre 1981,
Fuchs 2001, 2003c) so that density waves with smaller or greater wavelengths can
develop as well. Allowing for these side fringes of wavelengths and the
corresponding variations of the coefficient $X$ in equation (5) one can derive
for the distance range in the galactic disks spanned by the spiral arms a 
uniform value of $m$.

\section{Examples: NGC\,3198 and NGC\,2985}
I demonstrate the implications of the dynamical constraints on the decomposition
of the rotation curves with the examples NGC\,3198 and NGC\,2985. Both galaxies
show clear spiral structure and the velocity dispersions of the stars have been
measured in both galaxies (Bottema 1988, Gerssen 2000). The expected
multiplicity of spiral arms and the $Q$ parameter are shown in Fig.~3 for the
maximum disk and the submaximal disk models of NGC\,3198, respectively. The
maximum disk model predicts a two--armed spiral just as can be seen in the NIR
image of the galaxy in Fig.~2. The $Q$ parameter is about 1 in this model. This
avoids the the onset of fierce dynamical instabilities, but would imply that the
spiral structure grows very rapidly which would lead to a considerable dynamical
disk heating. However, the velocity dispersions have been measured by Bottema
(1988) in the B band and might be dominated by bright young stars with velocity
dispersions lower than that of the mass carrying stellar populations (Fuchs 
1999). Thus the $Q$ parameter is probably underestimated. The submaximal disk
model predicts a three--armed spiral and the $Q$ parameter is around 1.5 which
is probably underestimated again and thus seems to be too high.
 
\begin{center}
\hbox{
\epsfclipon
\epsfxsize=5.4cm
\epsffile{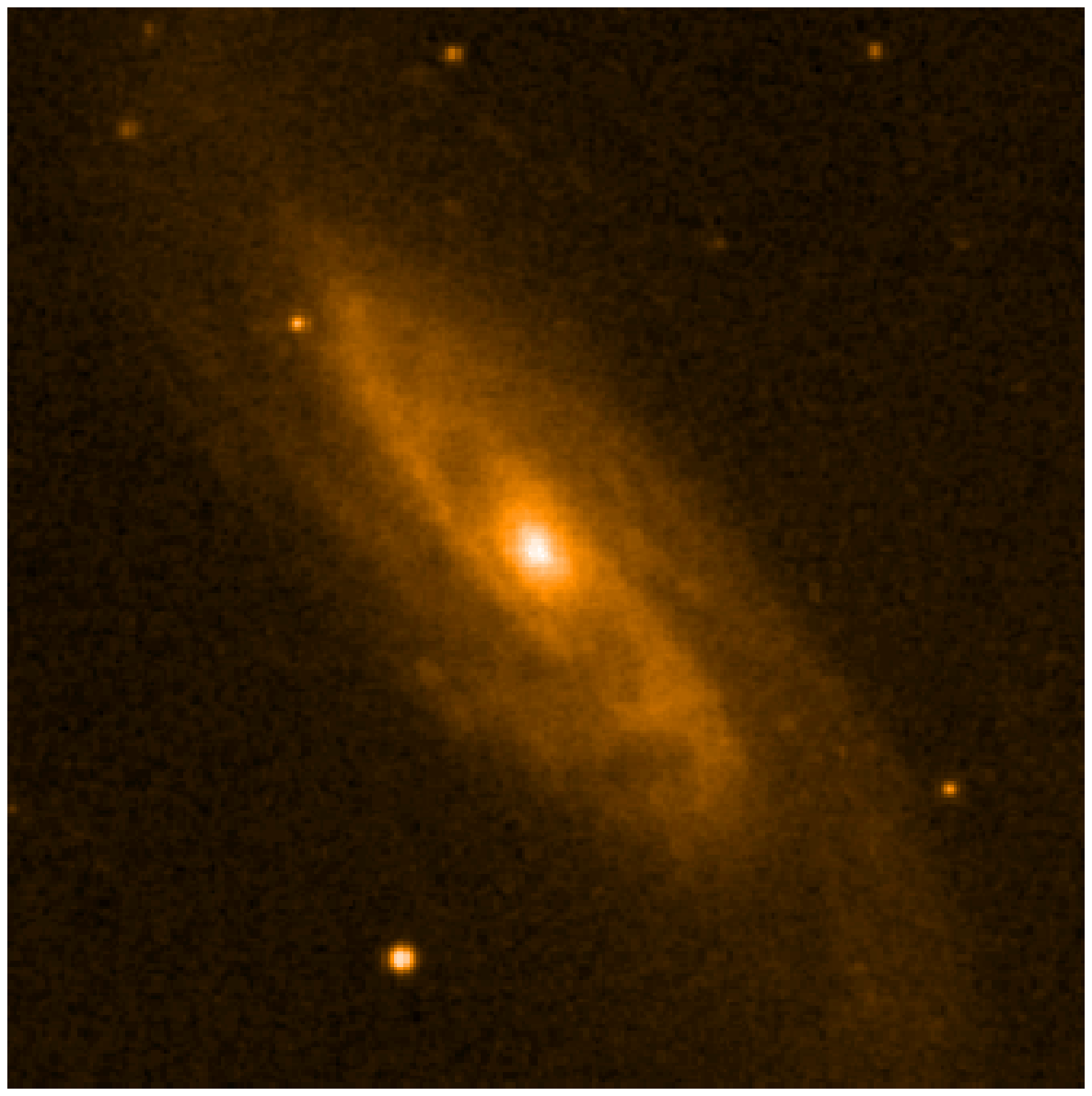}
\hspace{0.2cm}
\epsfclipon
\epsfxsize=5.4cm
\epsffile{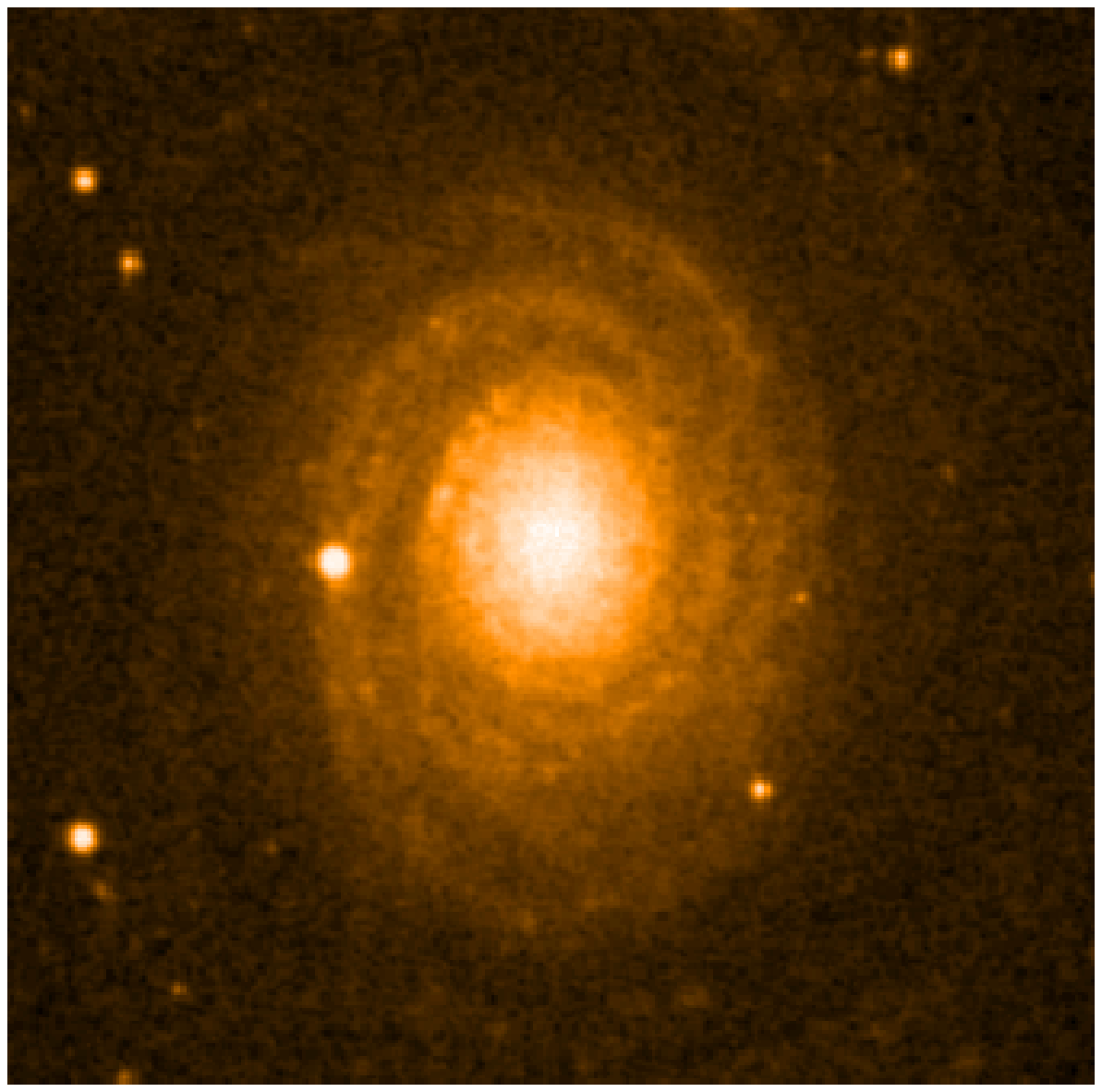}
}
\end{center}
\vspace{-0.5cm}
{\footnotesize Figure 2. Images of NGC\,3198 (left) and NGC\,2985 (right)
retrieved from the Digitized Sky Survey (ESO). The image sizes are 
5'$\times$5'.}

\begin{center}
\hbox{
\epsfclipon
\epsfxsize=5.4cm
\epsffile{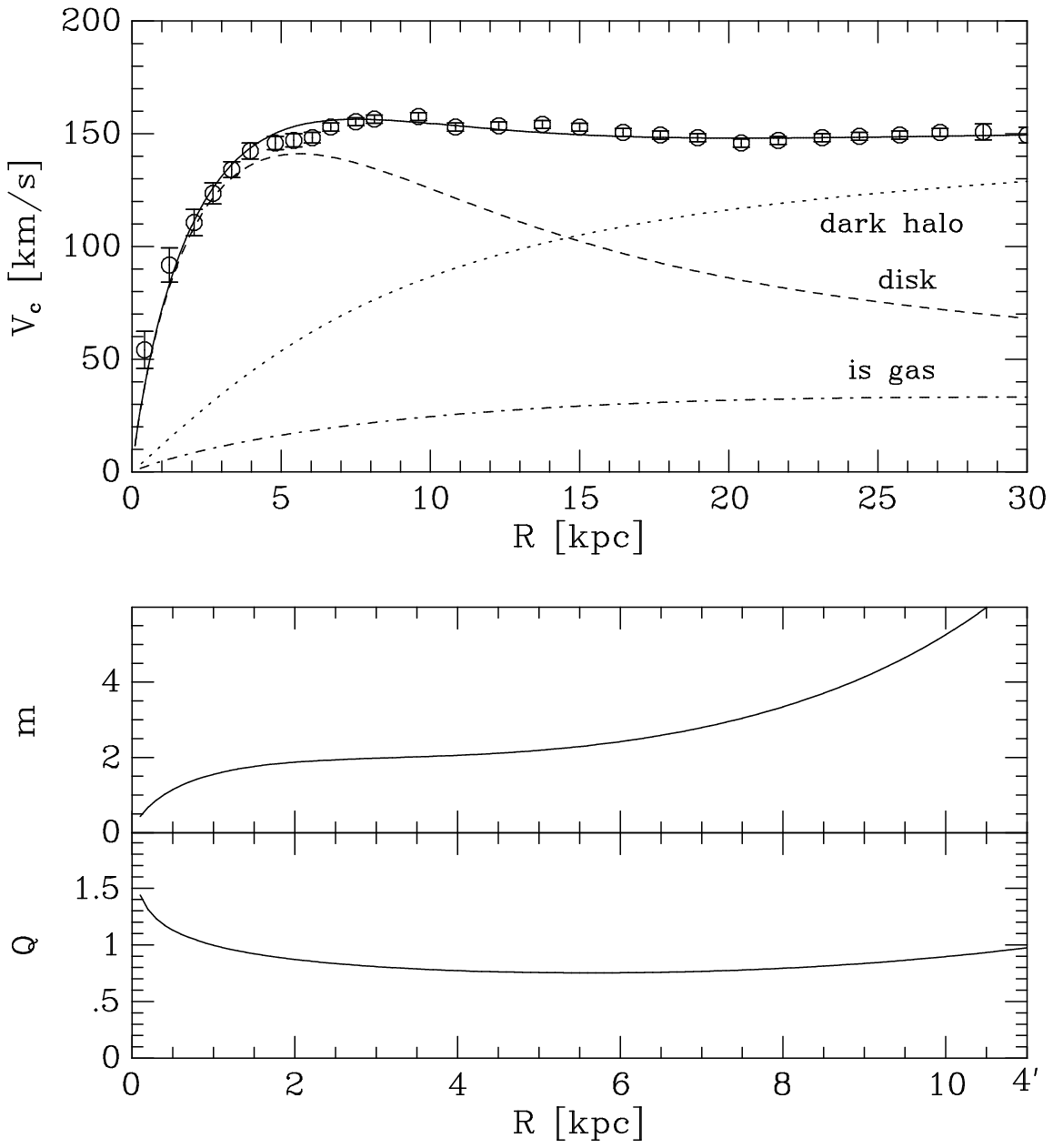}
\hspace{0.2cm}
\epsfclipon
\epsfxsize=5.4cm
\epsffile{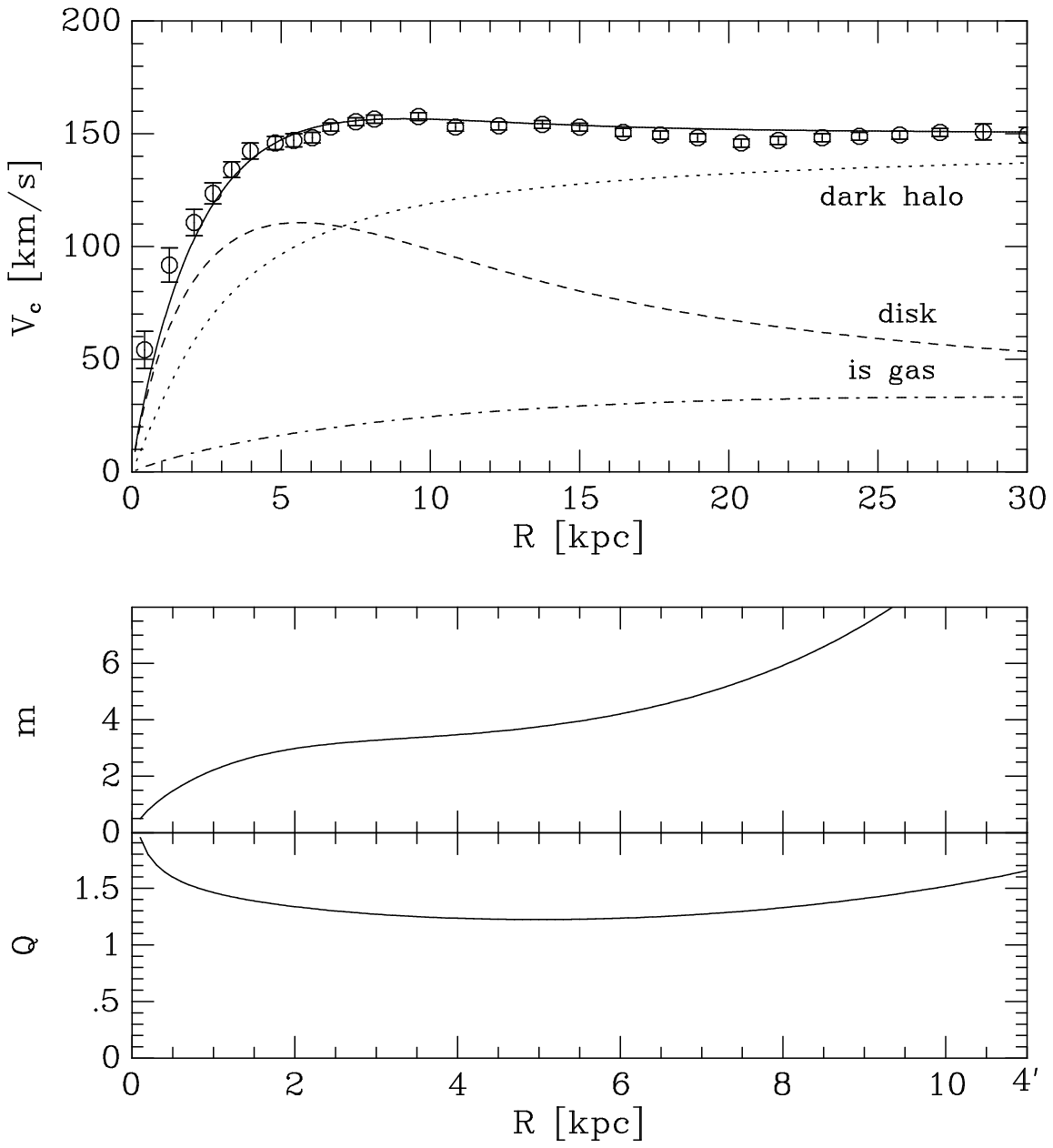}
}
\end{center}
\vspace{-0.5cm}
{\footnotesize Figure 3. Expected number of spiral arms and $Q$ parameter
according to the maximum disk and submaximal disk models of NGC\,3198 shown 
again in the upper panels. The radial scale is also indicated in arcmin.}
\vspace{0.2cm}

\noindent
I conclude from this that the maximum disk model is the more realistic disk
model. In a previous paper (Fuchs 1999) I have argued to the contrary.
However, that was judging from an optical image with many filaments which does
not reveal the major spiral arms as the NIR image and I revise my opinion
here. 

Fig.~4 shows the rotation curve of NGC\,2985 (Gerssen 2000) and the 
corresponding maximum
disk and submaximal disk models, respectively, which in both cases include a
bulge contribution in the inner parts. The velocity dispersions have been
measured by Gerssen (2000) in the I band and should allow a more reliable 
estimate of the $Q$ parameter than in the previous case. As can be seen from 
Fig.~4 both the expected multiplicity of spiral arms and the $Q$ parameter
indicate even clearer than in the case of NGC\,3198 that the maximum disk model
is the more realistic disk model. 

\begin{center}
\hbox{
\epsfclipon
\epsfxsize=5.4cm
\epsffile{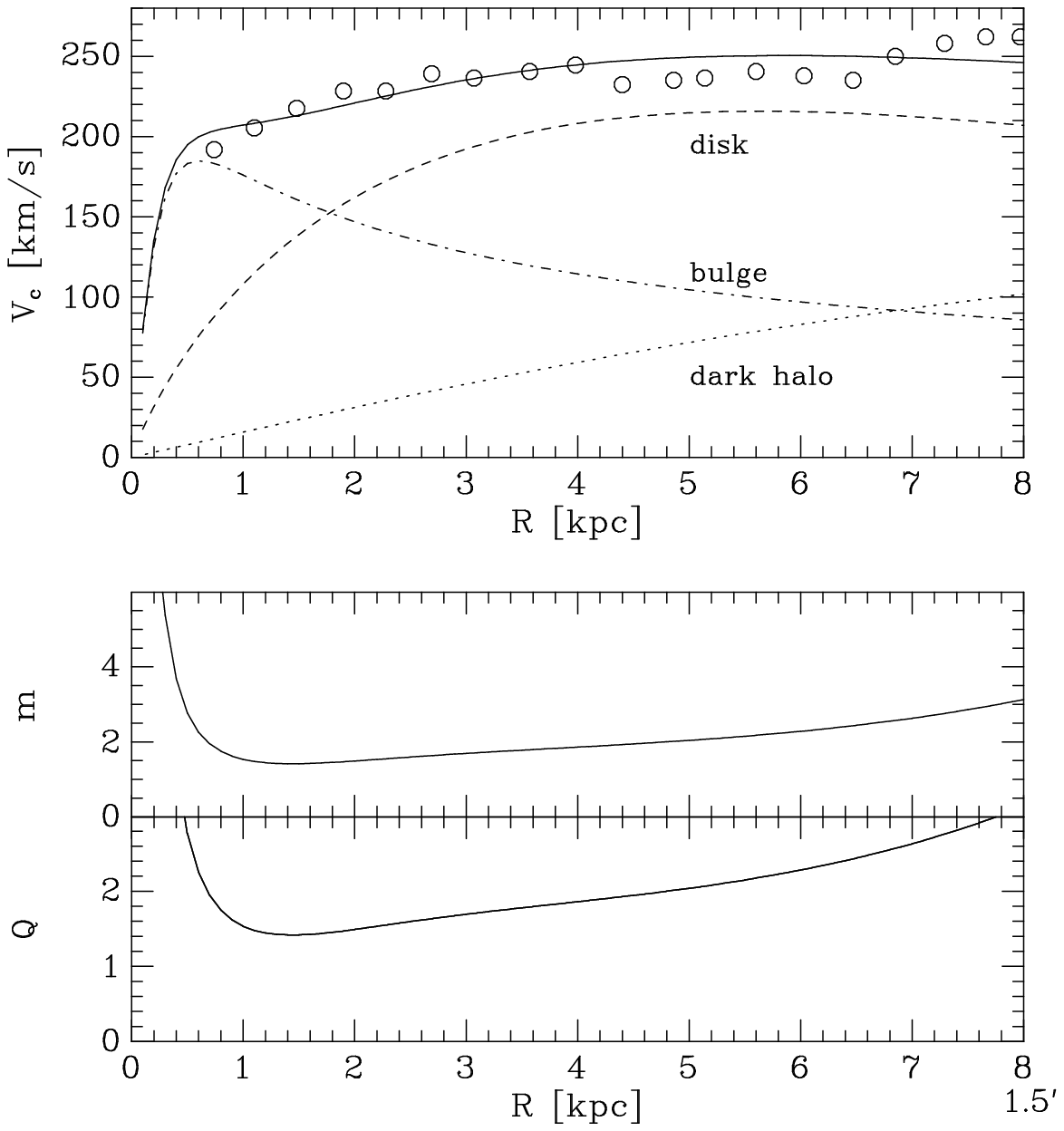}
\hspace{0.2cm}
\epsfclipon
\epsfxsize=5.4cm
\epsffile{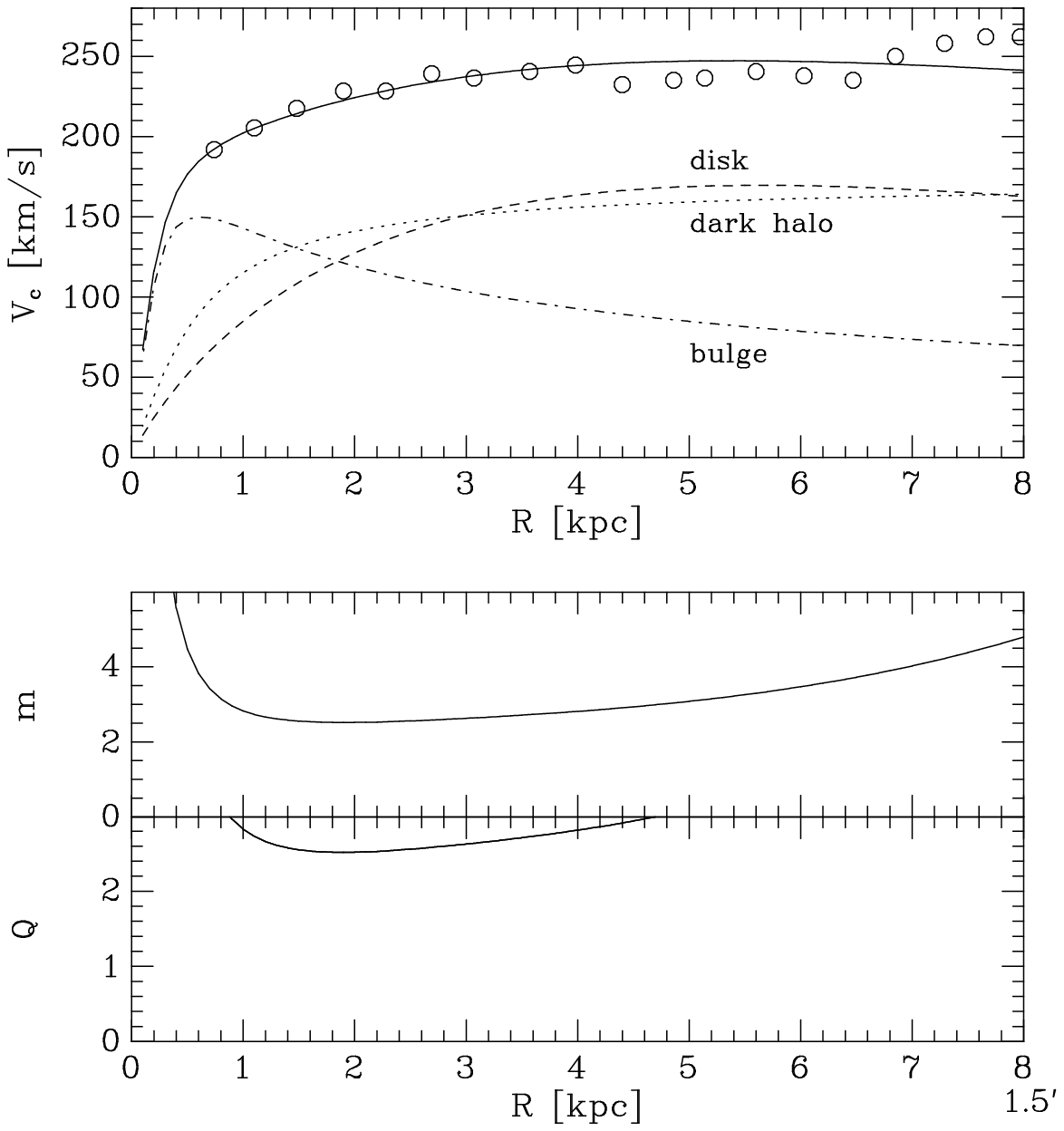}
}
\end{center}
\vspace{-0.5cm}
{\footnotesize Figure 4. Expected number of spiral arms and $Q$ parameter
according to the maximum disk and submaximal disk models of NGC\,2985 shown in
the upper panels. The radial scale is also indicated in arcmin.}

\section{Discussion}
Arguments for and against maximum disks have been discussed at length in the
literature and are reviewed in detail, for instance, by Bosma (1999) or Sellwood
(1999). One of the major consequences of maximum disks are the implied large
core radii of the dark halos. These challenge the contemporary theory of the
formation of galaxies according to CDM cosmology.

Obviously the dynamical constraints on the decomposition of rotation curves
must be tried out on a much larger data set than here in order to
test the maximum disk hypothesis. This can be easily done
with the density wave theory criterion by inspecting images of galaxies
for which rotation curves are available. I have, for instance, analyzed a set
of low surface brightness galaxies and found again indications for maximum
disks (Fuchs 2003a, b). However, velocity dispersions have been measured in
very few galaxies. Thus the $Q$ stability parameter criterion, which is an 
independent consistency check on the density wave theory criterion, can be 
applied only in deplorably few cases.

\section*{Acknowledgments}
The modelling of the rotation curve of NGC\,2985 was done in collaboration with
I. Arifyanto.


\begin{thebibliography}{0}
\bibitem{at} E. Athanassoula, A. Bosma, and S. Papaioannou, {\it A\&A}
 {\bf 179}, 23 (1987).
 
\bibitem{bo} A. Bosma, in {\it Galaxy Dynamics}, D. Merritt, J.A. Sellwood,
and M. Valluri (eds.), ASP Conf. Ser. {\bf 182}, 339 (1999).

\bibitem{bu} R. Bottema, {\it A\&A} {\bf 197}, 105 (1988).

\bibitem{br} A. H. Broeils, PhD thesis, Univ. of Groningen (1992).

\bibitem{fu} B. Fuchs, in {\it Galaxy Dynamics}, D. Merritt, J.A. Sellwood,
and  M. Valluri (eds.), ASP Conf. Ser. {\bf 182}, 365 (1999).

\bibitem{fuc} B. Fuchs, {\it A\&A} {\bf 368}, 107 (2001).

\bibitem{fuchs} B. Fuchs, in {\it Proc. Fourth Int. Conf. on Dark Matter in
Astro and Particle Physics}, H. V. Klapdor-Kleingrothaus and R. Viollier (eds.),
Springer, in press (2003a).

\bibitem{fuchsb} B. Fuchs, in {\it The Evolution of Galaxies. III--From Simple
Approaches to Self-consistent Models}, G. Hensler, G. Stasinska, S. Harfst, P.
Kroupa, and C. Theis (eds.), Kluwer, in press (2003b).

\bibitem{fuch} B. Fuchs, {\it A\&A} submitted (2003c).

\bibitem{fvl} B. Fuchs \& S. von Linden, {\it MNRAS} {\bf 294}, 513 (1998).

\bibitem{ger} J. Gerssen, PhD thesis, Univ. of Groningen (2000).

\bibitem{glb} P. Goldreich and D. Lynden-Bell, {\it MNRAS} {\bf 130}, 125
(1965).

\bibitem{jt} W. H. Julian \& A. Toomre, {\it ApJ} {\bf 146}, 810 (1966).

\bibitem{se} J. A. Sellwood, in {\it Galaxy Dynamics}, D. Merritt, J.A. 
Sellwood, and M. Valluri (eds.), ASP Conf. Ser. {\bf 182}, 351 (1999).

\bibitem{too} A. Toomre, in {\it Structure and Evolution of Normal Galaxies}, S. M.
Fall and D. Lynden-Bell (eds.), Cambridge Univ. Press, 111 (1981).

\end{thebibliography}
\end{document}